\documentclass{article}

\usepackage{PRIMEarxiv}

\usepackage[utf8]{inputenc} 
\usepackage[T1]{fontenc}    
\usepackage{hyperref}       
\usepackage{url}            
\usepackage{booktabs}       
\usepackage{amsfonts}       
\usepackage{nicefrac}       
\usepackage{microtype}      
\usepackage{lipsum}
\usepackage{fancyhdr}       
\usepackage{graphicx}       
\graphicspath{{media/}}     
\usepackage{caption}
\usepackage{subcaption}

\usepackage{multirow}
\usepackage{amsmath,amssymb}
\usepackage{bm}
\newcommand{\mi}{{\rm{i}}}
\newcommand{\rrA}{{\bm{r}}_{\rm A}}
\newcommand{\rr}{{\bm{r}}}

\title{Onsager's Real Cavity model near solid interfaces
}

\author{
  Johannes Fiedler \\
  Department of Physics and Technology\\
  University of Bergen\\
  All\'egaten 55, 5007 Bergen, Norway.\\
  \texttt{johannes.fiedler@uib.no} \\
   \And
  Drew F. Parsons \\
  Department of Chemical and Geological Sciences\\
  University of Cagliari\\
  Cittadella Universitaria, 09042 Monserrato, CA, Italy.\\
  \texttt{drew.parsons@unica.it} \\
}

\begin{document}
\maketitle

\begin{abstract}
We develop an extended Onsager real-cavity framework to describe the Casimir--Polder interaction of small molecules dissolved in dielectric liquids near planar interfaces. By analytically resolving the geometry of the cavity opening, we derive closed-form expressions that capture the modification of the interaction as the molecule approaches a surface and connect smoothly to the asymptotic medium-assisted limit. 

Using experimentally established dielectric functions for water, propanol, and PTFE together with accurate molecular polarisabilities for O$_2$ and N$_2$, we compute the full distance-dependent potential for representative molecule--liquid--surface combinations. The results reveal how local-field screening, cavity geometry, and material response jointly determine both the magnitude and shape of the interaction, including the characteristic transition between open-cavity (\(z\lesssim z_{\rm C}\)) and closed-cavity (\(z\gtrsim z_{\rm C}\)) regimes.

Beyond providing quantitative predictions, the framework offers an analytically transparent decomposition of dispersion forces in liquids, enabling a direct identification of the underlying physical contributions and an efficient exploration of parameter dependencies across different systems. The approach thus provides a useful baseline for interpreting dispersion interactions in complex environments within a continuum, local-field corrected description.
\end{abstract}

\section{Introduction}

Van der Waals forces are fundamental interactions between neutral, polarisable particles~\cite{London1937,VanderWaals1873,London1930}, crucial for cohesion in materials and for biological adhesion phenomena, such as geckos walking on smooth surfaces~\cite{Autumn2000}. 
These dispersion forces are increasingly relevant in nanoscale technologies~\cite{DelRio2005,Fiedler_2022,Akhundzada2022,Fiedler2024,PhysRevApplied.13.014025,Osestad2025} and have been thoroughly studied both experimentally~\cite{Sukenik1993CPMeasurement,Lamoreaux1997Casimir,Mohideen1998Casimir,Munday2009Repulsive,Grisenti1999,Arndt1999,Juffmann2012,Brand2015} and theoretically~\cite{Casimir48,Craig1984MQED,Salam2016vdW,pitaevskii,Scheel2008,Buhmann12a}.

Dispersion interactions emerge from correlated quantum fluctuations of charges and electromagnetic fields and can be described within quantum electrodynamics as fluctuation-induced interactions between neutral bodies. 
Depending on the separation regime and electromagnetic environment, this leads to different limiting forms commonly referred to as London--van der Waals~\cite{London1930}, Casimir--Polder~\cite{Casimir48}, or Casimir interactions~\cite{Casimir482}. 
While these interactions are among the weakest fundamental forces, they nevertheless govern a wide range of physical, chemical, and biological processes.

Despite their origin in quantum electrodynamic fluctuations~\cite{D2CP03349F}, dispersion forces—such as the Casimir~\cite{Casimir482}, Casimir--Polder~\cite{Casimir48}, and London--van der Waals~\cite{London1930} forces—are among the weakest forces in nature. They arise from correlated ground-state fluctuations of the electromagnetic field. While classical approaches often model these forces in vacuum~\cite{Casimir48,London1937,McLachlan387}, real-world systems such as colloids, proteins, or nanoparticles typically reside in polar media like water or organic solvents~\cite{MILONNI1994217}.

In addition to modifying dispersion forces, electromagnetic environments can also alter resonant energy-transfer processes and induce discriminatory interactions between chiral molecules. Recent macroscopic QED studies have demonstrated that dielectric and chiral surroundings can qualitatively modify intermolecular transfer rates and interaction pathways through local-field and medium-assisted effects~\cite{Franz2023RET,Franz2024ChiralRET}. These developments further illustrate the importance of consistently incorporating environmental response into fluctuation-induced interaction theories.

The presence of a surrounding medium alters both the range and magnitude of dispersion interactions~\cite{Ether_2015,LOSKILL2012107}, often invalidating simple vacuum-based asymptotics~\cite{PhysRevLett.104.070404}. The interplay between retardation, thermal fluctuations, and dielectric screening can result in nonmonotonic potentials and even repulsive forces~\cite{PhysRevLett.106.064501,PhysRevA.81.062502,PhysRevB.101.235424}. Such features enable novel trapping mechanisms~\cite{Zhao984,DouPhysRevB.89.201407,doi:10.1021/acs.jpcc.8b02351,MathiasPhysRevB2018} without the need for external stabilising forces.

While various microscopic approaches (e.g., molecular dynamics~\cite{doi:10.1021/jp012750g,doi:10.1021/jp020242g,doi:10.1021/jp065191s,doi:10.1021/jp804811u}) provide detailed insight into solvent structure and interfacial phenomena~\cite{annurev:/content/journals/10.1146/annurev.physchem.57.032905.104609,doi:10.1021/cr0403741,Levin_2014}, they typically yield numerical results whose physical interpretation and parameter dependence can be difficult to disentangle. Polarisable continuum models (PCM) based on dielectric theory~\cite{10.1063/1.1643727}, such as those by Levin et al.~\cite{PhysRevLett.103.257802}, offer a complementary perspective~\cite{doi:10.1063/5.0037629,D3NR05396B,DuignanParsonsNinham2014surfaceTension,DuignanParsonsNinham2015,doi:10.1021/acsphyschemau.6c00007}. The PCM model and the related COSMO model \cite{Klamt2018} have been implemented in quantum chemical software to enabling quantification of the impact of solvent polarisation on the electron structure of a molecule \cite{CramerTruhlar2008,KlamtEtAl2009}. PCM approaches are typically better aligned with the underlying physics than purely microscopic configurational models~\cite{doi:10.1021/acs.jpca.5c03229,10.1063/5.0216468}, especially when addressing macroscopic observables that inherently involve long-range electromagnetic response, such as optical or dielectric properties. PCM models suggest that cavity formation energies and dielectric response can drive adsorption or exclusion at interfaces. One shortcoming of PCM models is that the dielectric medium is typically  characterised solely by the static dielectric constant of the medium. This may be addressed in part by including a cavity energy term in the total energy alongside dispersion interactions \cite{PhysRevLett.103.257802,DuignanParsonsNinham2014surfaceTension}. The cavity energy may be taken as zeroth order term in molecule-solvent interactions. Hydrogen bonding between water solvent and a solute molecule may be considered as a higher order correction, or treated by taking the molecule as a hydrated complex including explicit hydration waters \cite{ParsonsCarucciSalis2022} (inside the cavity) . But the cavity energy alone neglects molecule-solvent dispersion interactions that  arise from the full dielectric spectrum employed in macroscopic QED. Taken together the complete  model may describe ion solvation energies \cite{DuignanParsonsNinham2013} or surface tension increments \cite{DuignanParsonsNinham2014surfaceTension,DuignanParsonsNinham2015}. The latter work considered the approach of ions at the air-water interface, with cavity and dispersion energies evaluated as a simple fraction of cavity area remaining after contact with air. In this work we develop a similar but more exact theory for the dispersion energy in the interfacial region between a solvent and a solid dielectric material.

Beyond their conceptual simplicity, continuum descriptions of the electromagnetic environment remain a central and actively used modelling framework even at nanometre length scales. Modern continuum approaches are routinely employed to describe dielectric screening, dispersion forces, solvation energies, and electromagnetic response in complex soft-matter and condensed-phase systems, provided that local-field effects and appropriate microscopic boundary conditions are consistently incorporated. In particular, the introduction of an effective excluded-volume region around the solute allows one to separate short-range molecular structure from long-range electromagnetic response, thereby enabling a controlled matching between microscopic physics and macroscopic electrodynamics. This strategy underlies a wide range of contemporary applications, including polarisable continuum models in quantum chemistry, nanoscale dispersion interactions near interfaces, and effective medium descriptions of molecular and nanostructured systems, where fully microscopic treatments remain computationally demanding while continuum theories retain predictive power when properly formulated~\cite{PCMbook,water_ano}.

In this work, we extend this continuum framework to study the Casimir--Polder interaction between a solvated particle (treated as a point dipole) and a planar interface within a polarisable continuum model. We focus on the regime in which the particle--surface separation becomes comparable to the characteristic cavity size, such that the effective excluded-volume region surrounding the molecule is distorted by the interface and partially opens towards it. This geometric modification leads to a non-trivial restructuring of the local electromagnetic environment and, consequently, of the dispersion interaction.

By explicitly resolving the geometry of the cavity opening, we derive closed-form analytic expressions that decompose the interaction into distinct geometric contributions and connect smoothly to the asymptotic medium-assisted Casimir--Polder limit at larger separations. This analytic structure provides direct physical insight into how local-field screening, cavity geometry, and material response combine to shape dispersion forces in liquids. At the same time, it enables efficient exploration of parameter dependencies across different solvents, molecular species, and surfaces, offering a transparent baseline for interpreting more detailed microscopic calculations.

The present description is based on a continuum dielectric response and is therefore expected to be quantitatively reliable once the particle--surface separation exceeds a few molecular or lattice spacings. At smaller separations, microscopic structure, electronic overlap, and short-range interactions become increasingly important, and the continuum description should be regarded as an effective extrapolation. Within this well-defined regime, however, the approach captures the essential interplay between geometry and electromagnetic response that governs dispersion interactions in solvated systems.

\section{Modelling}
The dispersion interaction between a neutral, polarisable particle at position $\rrA$ with polarisability ${\bm{\alpha}}(\omega)$ and a macroscopic body at temperature $T$ is determined by the Casimir--Polder interaction~\cite{Scheel2008}
\begin{equation}
   U_{\rm CP}(\rrA) = \mu_0 k_{\rm B}T \sum_{n=0}^\infty {}'\xi_n^2 \operatorname{Tr} \left[ {\bm{\alpha}}(\mi\xi_n)\cdot {\bf{G}}(\rrA,\rrA,\mi \xi_n)\right] \,,\label{eq:UCP}
\end{equation}
with the vacuum permeability $\mu_0$ and the Boltzmann constant $k_{\rm B}$. At non-zero temperature it is convenient to evaluate Eq.~(\ref{eq:UCP}) on the imaginary frequency axis at the Matsubara frequencies $\hbar\xi_n = 2\pi nk_{\rm B} T$, where $\hbar$ is the reduced Planck constant. The scattering Green tensor ${\bf{G}}(\rrA,\rrA,\mi\xi)$ encodes the linear electromagnetic response of the surrounding bodies; equivalently, one may rewrite Eq.~(\ref{eq:UCP}) as an integral over real frequencies, where both propagating and evanescent contributions appear.
The primed sum in Eq.~(\ref{eq:UCP}) means that the $n=0$ term of the sum is weighted by a factor $1/2$,
\begin{equation}
    \sum_{n=0}^\infty{}' a_n = \frac{1}{2}a_0 +  \sum_{n=1}^\infty a_n \,.
\end{equation}

\subsection{Solvent effects}
We consider a particle in a liquid with dielectric function $\varepsilon_{\rm L}(\omega)$ located in front of a dielectric planar surface with dielectric function $\varepsilon_{\rm S}(\omega)$. In the non-retarded short-distance regime (particle--surface separation much smaller than the relevant electromagnetic wavelengths), the interaction reduces to the familiar \(C_3\) form
\begin{equation}
    U_{\rm CP}(\rrA) = -\frac{C_3}{z_{\rm A}^3} \,. \label{eq:UPCC3}
\end{equation}
This representation should not be confused with the general-distance result of Eq.~(\ref{eq:UCP}); it follows from the planar Green tensor in the non-retarded limit. This potential is derived from the scattering Green function for a planar two-layer system~\cite{doi:10.1021/acs.jpca.7b10159}
\begin{eqnarray}
    {\bf{G}}(\rrA,\rrA,\omega) = \frac{c^2}{32\pi\varepsilon_{\rm L}(\omega)\omega^2 z_{\rm A}^3}\frac{\varepsilon_{\rm S}(\omega)-\varepsilon_{\rm L}(\omega)}{\varepsilon_{\rm S}(\omega)+\varepsilon_{\rm L}(\omega)}\begin{pmatrix}1 & 0 & 0 \\ 0 & 1 & 0 \\ 0&0&2
    \end{pmatrix}\,, \label{eq:Gplanar}
\end{eqnarray}
with the speed of light $c$. Inserting the planar Green function~(\ref{eq:Gplanar}) into the Casimir--Polder formulation of ~(\ref{eq:UCP}) determines the dispersion coefficient $C_3$,
\begin{equation}
    C_3 = \frac{k_{\rm B}T}{8\pi\varepsilon_0}\sum_{n=0}^\infty {}' \frac{\alpha(\mi\xi_n)}{\varepsilon_{\rm L}(\mi\xi)}\frac{\varepsilon_{\rm S}(\omega)-\varepsilon_{\rm L}(\omega)}{\varepsilon_{\rm S}(\omega)+\varepsilon_{\rm L}(\omega)}\,,
\end{equation}
with  vacuum permittivity $\varepsilon_0 = \left(\mu_0c^2\right)^{-1}$. It can be seen that the permittivity of the surrounding medium screens the medium-assisted $C_3$ coefficient $\varepsilon_{\rm L}(\mi\xi)$ compared to the same situation in vacuum ($\varepsilon_{\rm L}(\mi\xi)=1$). Due to the electrodynamical (frequency-dependent) screening, the van der Waals interaction between two particles in water is effectively suppressed by a factor of two. In contrast, the Coulomb force is suppressed by a factor of 80.~\cite{doi:10.1063/5.0037629} 

Beyond dielectric screening, local-field effects are commonly modelled by introducing an effective excluded-volume region around the solute (``cavity'') that accounts for the microscopic depletion of solvent close to the solute and provides a well-defined boundary for electromagnetic matching in continuum theory. The effect is usually considered via the excess or effective polarisability models,~\cite{doi:10.1021/acs.jpca.7b10159} which modifies the polarisabilities to include local-field corrections for the photon propagation. The simplest model considers a point-like particle in the centre of a spherical vacuum cavity with radius $R_{\rm C} = \sqrt[3]{3V_{\rm C}/(4\pi)}$ known as Onsager's real cavity model~\cite{doi:10.1021/acs.jpca.7b10159,doi:10.1021/ja01299a050}
\begin{equation}
    \alpha(\mi\xi) \mapsto \left(\frac{3\varepsilon_{\rm L}(\mi\xi)}{1+2\varepsilon_{\rm L}(\mi\xi)}\right)^2 \alpha(\mi\xi) \,,\label{eq:Onsager}
\end{equation}
which is the square of the zeroth-order Mie transmission coefficient, which includes the inwards and outwards propagation of the photon through the interface. However, this model is restricted to the scenario that the same media surrounds the particle in all directions. In particular, the continuum description is expected to be appropriate only for separations larger than molecular length scales; below that, microscopic structure, exchange/overlap, and chemical specificity enter.

\subsection{The open cavity}\label{sec:OpenCav}
Onsager's real cavity model, introduced in the previous section, is restricted to particles homogeneously surrounded by a medium. This model will fail when the particle approaches a solid surface, where the medium will be displaced. We applied the Born series expansion for the scattering Green function to model this scenario.~\cite{Scheel2008,Buhmann12b} The Born/Hamaker representation employed below is explicitly a weak-scattering, non-retarded approximation; it is used here to obtain analytic geometry-resolved expressions and is subsequently matched to the exact planar \(C_3\) limit (rescaling).
This approach separates the scattering processes according to the number of scattering events. Its first order reads~\cite{PhysRevX.4.011029}
\begin{eqnarray}
   {\bf{G}}({\bm{r}}_{\rm A},{\bm{r}}_{\rm A},\omega) = \frac{\omega^2}{c^2}\int\limits_{\rm \mathbb{R}^3} \frac{\mathrm d^3 s\chi({\bm{s}},\omega)}{1+\chi({\bm{s}},\omega)/3}{\bf{R}}({\bm{r}}_{\rm A},{\bm{s}},\omega)\cdot {\bf{R}}({\bm{s}},{\bm{r}}_{\rm A},\omega)\,,\label{eq:Green}
\end{eqnarray}
with the spatially-dependent susceptibility of the surrounding media $\chi({\bm{s}},\omega) = \varepsilon({\bm{s}},\omega) -1$ and the regular part of the free-space Green function~\cite{PhysRevX.4.011029}
\begin{eqnarray}
    {\bf{R}}({\bm{r}},{\bm{r}}',\omega)  = \frac{q}{4\pi }\left[f\left(\frac{1}{q\varrho}\right)\mathbb{I}- g\left(\frac{1}{q\varrho}\right)\frac{{\bm{\varrho}}\otimes{\bm{\varrho}}}{\varrho^2}\right]\mathrm e^{{\rm i}q \varrho },\label{eq:RR}
\end{eqnarray}
 with the three-dimensional unit matrix $\mathbb{I}$, the relative coordinate ${\bm{\varrho}}={\bm{r}}-{\bm{r}}'$, its absolute value $\varrho=\left|{\bm{\varrho}}\right|$, absolute value of the wave vector $q=\omega/c$ and the functions $f(x) = x+\mi x^2-x^3$ and $g(x) = x+3\mi x^2-3x^3$. By assuming the particle-surface separations $d$ to be smaller than the wavelength of the dominant atomic transition, $d\ll \omega_{\rm max}/c$, the non-retarded limit can be applied to lead to
\begin{eqnarray}
\operatorname{Tr}\left[{\bf{R}}({\bm{r}},{\bm{s}},\mi\xi)\cdot{\bf{R}}({\bm{s}},{\bm{r}},\mi\xi)\right] = \frac{3c^4}{8\pi^2\xi^4\varrho^6 }\,.
\end{eqnarray}
Thus, to first order, the Casimir--Polder potential within an arbitrary environment can be written via the Hamaker approach~\cite{Hamaker} \begin{eqnarray}
U_{\rm CP}(\rrA) = -\frac{9k_{\rm B}T}{8\pi^2\varepsilon_0}\sum_{n=0}^\infty {}' \alpha(\mi\xi_n) \int\limits_{\mathbb{R}^3} \mathrm d^3 s\frac{\varepsilon({\bm{s}},\omega)-1}{\varepsilon({\bm{s}},\omega)+2} \frac{1}{\left|\rrA -{\bm{s}}\right|^6} \,,\label{eq:UCPnonrenom}
\end{eqnarray}
summing (integrating) the pairwise interactions between a particle at position $\rrA$ and a volume element at position ${\bm{s}}$. Equation~(\ref{eq:UCPnonrenom}) is a non-retarded, pairwise-additive approximation and should be interpreted as such. Hence, the Casimir--Polder interaction is determined by the summation (integration) of van-der-Waals interactions. This model is convenient to introduce the geometry of a system beyond the well-known analytical systems (planar, cylindrical, and spherical multilayered systems)~\cite{Scheel2008,Buhmann12b} by considering weakly responding materials. In Eq.~\eqref{eq:UCPnonrenom}, the Mie reflection coefficient, $(\varepsilon-1)/(\varepsilon+2)$, can be seen, which will generate a deviation from the well-known Casimir--Polder scenario in front of an infinite half-plane, which requires the Fresnel-coefficient, $(\varepsilon-1)/(\varepsilon+1)$. Thus, to ensure reproducing the correct limit for a planar interface, this Hamaker approach needs to be rescaled, in analogy to Refs.~\cite{Brand2015,Fiedler_2022}, and thus the Casimir--Polder interaction reads
\begin{equation}
    U_{\rm CP}(\rrA) =-\frac{6}{\pi} \int\mathrm d^3 s \frac{C_3({\bm{s}})}{\left|\rrA -{\bm{s}}\right|^6} \,,\label{eq:UCPapp}
\end{equation}
with the $C_3$-coefficient of the volume element at position ${\bm{s}}$
\begin{equation}
  C_3({\bm{s}})=\frac{k_{\rm B}T}{8\pi\varepsilon_0}\sum_{n=0}^\infty {}' \alpha(\mi\xi_n) \frac{\varepsilon({\bm{s}},\mi\xi_n)-1}{\varepsilon({\bm{s}},\mi\xi_n)+1}\,.
 \end{equation}

We note that the present theory is evaluated in the non-retarded regime, where the interaction continuously connects to the familiar London dispersion interaction scaling. In this limit, the Casimir--Polder formalism reduces to the short-range van der Waals interaction between fluctuating dipoles and surfaces. The distinction is therefore primarily one of physical regime and theoretical formulation rather than different underlying mechanisms.

By contrast, Lennard--Jones potentials provide an empirical representation of dispersion interactions alongside short-range repulsion that represents  overlap (the quantum mechanical exchange energy) of the electron clouds of interacting molecules \cite{LennardJones1931}.
Unlike Lennard--Jones potentials, which are typically parameterised from experiment or electronic-structure calculations \cite{MatyushovSchmid1997,SchwerdtfegerWales2024}, the macroscopic QED framework provides a microscopic field-theoretical description that naturally incorporates dielectric environments, retardation, and local-field effects through the electromagnetic Green tensor and frequency-dependent response functions.

\begin{figure}[t]
    \centering
    \includegraphics[width=0.6\columnwidth]{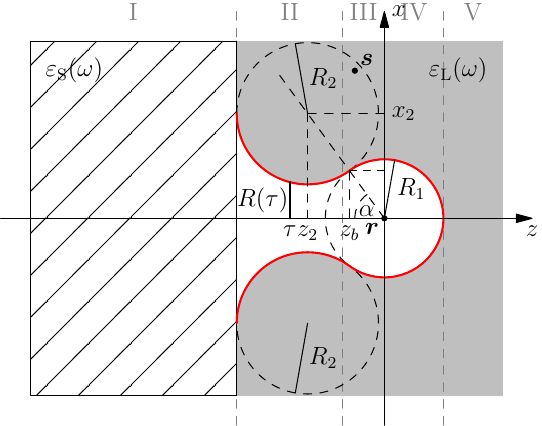}
    \caption{Cross-section through the considered system: a point-particle located at $\rr$ embedded in a medium (grey area) with dielectric function $\varepsilon_{\rm L}(\omega)$ creates a vacuum bubble with radius $R_1$ surrounding it. Close to a dielectric interface $\varepsilon_{\rm S}(\omega)$ (hatched area) at a distance $z$, the particle will displace the surrounding media and open the cavity. The single medium particles occupy a spherical volume with a radius $R_2$, leading to the opened cavity illustrated in white, bounded by a profile (red line). To describe the profile mathematically,  the centre of the circle of the medium particles is labelled with the coordinates $(z_2,x_2)$, and the touching points are in the plane $z=z_b$. The three-dimensional scenario is a body of rotation around the $z$-axis. To describe the Casimir--Polder interaction between the particle and the surrounding material, the van der Waals interaction must be integrated over the entire volume, split into five regions I--V.}
    \label{fig:system}
\end{figure}

To apply the generalised Hamaker approach~\eqref{eq:UCPapp} to the open cavity, the integration volumes need to be specified. Following the point-particle assumption, the resulting cavities have spherical shapes with the molecule's cavity radius $R_1$ and the radius of the occupation volume of a solvent molecule $R_2$. The resulting geometric arrangement of the solvent around the molecule will be rotationally symmetric according to the surface's normal as depicted in Fig.~\ref{fig:system}. The particle is located in the vacuum bubble (white area) with vanishing contrast; thus, the remaining volume integral has to be considered with respect to the solid surface (hatched area) and the surrounding liquid material (grey area), following the scenario described by Duignan et al.\cite{DuignanParsonsNinham2014surfaceTension}. Transforming the integral~\eqref{eq:UCPapp} into cylindrical coordinates ($\mathrm d^3 s = \varrho \mathrm d \varrho \mathrm d\varphi \mathrm d z$), the angular integral can be ruled out, and the radial boundary between the vacuum and liquid phase can be described by
\begin{equation}
    R(\tau;z) = \begin{cases} \sqrt{R_1^2-{\tau}^2} \,, & \text{if} \,\tau<z_b \,,\\
    x_2-\sqrt{R_2^2-\left(\tau-z_2 \right)^2} & \text{if} \,z_b \le \tau \le z 
    \end{cases}\,,
\end{equation}
with the boundary point $z_b = R_1 (z-R_2)/(R_1+R_2)$. The total geometry is described by both radii $R_1$ and $R_2$ and the molecule-surface-distance. Via the angle of the boundary point, $\cos\alpha = z_b/R_1$, the central point of the solvent's occupation volume can be derived $(z_2,x_2) = (R_1+R_2)(\cos\alpha,\sin\alpha)$.

To estimate the Casimir--Polder potential for this system according to Eq.~(\ref{eq:UCPapp}), the van der Waals interaction has to be integrated over the entire space, which reduces in the considered scenario to the integral over the hatched and grey areas in Fig.~\ref{fig:system}. Due to the linearity of the integral, the Casimir--Polder potential can be split into contributions from five regions, marked in Fig.~\ref{fig:system} with Roman numbering,
\begin{equation}
    U^{\rm OC}_{\rm CP}(z) = U_{\rm CP}^{\rm I} + U_{\rm CP}^{\rm II}(z) + U_{\rm CP}^{\rm III}(z) + U_{\rm CP}^{\rm IV+V}(z)\,.\label{eq:UCPopencav}
\end{equation}
The last two regions provide constant energy because they are independent of the separation between the particle and the interface, determined only by the cavity radius $R_1$,
\begin{equation}
    U_{\rm CP}^{\rm IV+V} = -\frac{4C_3^{\rm L}}{R_1^3} \,.\label{eq:UCPI+II}
\end{equation}
In the third region, the Casimir--Polder potential reads
\begin{equation}
    U_{\rm CP}^{\rm III}(z) = -\frac{3C_3^{\rm L}}{R_1^3}\frac{z-R_2}{R_1+R_2}\,.
\end{equation}
In the cavity-opening region II, the potential reads
\begin{equation}
    U_{\rm CP}^{\rm II}(z) = -\frac{12 C_3^{\rm L}R_2}{\pi R_1^3\left(R_1 + 2R_2\right)^3}\frac{\Lambda_{\rm N}}{\Lambda_{\rm D}}\,,
\end{equation}
where the expressions for $\Lambda_{\rm N}$ and $\Lambda_{\rm D}$ can be found in App.~\ref{app:long}. Finally, the interaction between the dissolved particle and the interface in region I
\begin{eqnarray}
    U_{\rm CP}^{\rm I} (z) = -\frac{C_3^{\rm S}}{{z}^3}\,,
\end{eqnarray}
with the $C_3$ coefficients for the \textit{l}iquid ($\rm{X} =\rm{L}$) and \textit{s}olid ($\rm{X} =\rm{S}$)
\begin{equation}
  C_3^{\rm X}=\frac{k_{\rm B}T}{8\pi\varepsilon_0}\sum_{n=0}^\infty {}' \alpha(\mi\xi_n) \frac{\varepsilon^{\rm X}(\mi\xi_n)-1}{\varepsilon^{\rm X}(\mi\xi_n)+1}\,. \label{eq:C3free}
 \end{equation}

\subsection{Rescaling corrections to the Hamaker approach}
The Casimir--Polder potential for the open Onsager cavity~\eqref{eq:UCPopencav} is based on the Hamaker approach~\eqref{eq:Green}, as the pairwise summation of the interaction. Thus, it captures only the leading order of the interaction~\cite{Brand2015,Fiedler_2022}. The Onsager's real cavity model considers a point particle embedded in a vacuum bubble, leading to the Casimir--Polder potential of the well-known $r^{-3}$ form, where the polarisability of the particle has to be replaced by the excess polarisability~\eqref{eq:Onsager}. This model requires a closed cavity surrounding the particle. The model, introduced in Sec.~\ref{sec:OpenCav}, for the open cavity closes at distance $z_{\rm C}= \sqrt{R_1\left(R_1-2R_2\right)}+R_2$, meaning that the validity of the model is restricted to particle-surface separations below this threshold, $z<z_{\rm C}$. However, for larger separations, the potential should continuously transition into the long-range Casimir--Polder potential. Thus, the potential can be written as
\begin{equation}
    U_{\rm CP}(z) = \begin{cases}
        \lambda U_{\rm CP}^{\rm OC}(z) \,,& \text{for}\, z< z_{\rm C}\\
        -\frac{\tilde{C}_3}{z^3}\,, &\text{for} \, z\ge z_{\rm C}
    \end{cases}\,,
\end{equation}
where the introduced scaling constant $\lambda$ ensures the potential's continuity and corrects the pairwise summation approximation. It is fixed uniquely by continuity at \(z=z_{\rm C}\),
\begin{equation}
\lambda = \frac{-\tilde{C}_3/z_{\rm C}^3}{U_{\rm CP}^{\rm OC}(z_{\rm C})}\,,
\end{equation}
and therefore depends only on the material parameters entering \(C_3^{\rm L}\), \(C_3^{\rm S}\), and \(\tilde{C}_3\), and on the geometry through \(R_1,R_2\). In the numerical examples below, we find \(\lambda\) to be of order unity, and variations of \(\lambda\) within a few per cent merely shift the near-\(z_{\rm C}\) crossover without changing the qualitative region hierarchy.

\subsection{The contact limit}
At contact, i.e. for particle--surface separation $z\to 0$, the Casimir--Polder potential remains finite. In this limit, the relevant electromagnetic response can be viewed as that of a half-open (hemispherical) cavity in direct contact with the solid interface, such that the contributions from Regions~II and~III vanish and only Regions~I, IV, and~V remain. Regions~IV+V yield the constant offset potential \eqref{eq:UCPI+II}. In the contact limit $z\to0$, the geometry of Region~I reduces to the standard problem of an atom in vacuum in front of a planar dielectric interface. In this situation, the Casimir--Polder interaction is fully determined by the scattering part of the planar Green tensor, while the free-space contribution corresponds to the Lamb shift and does not enter the interaction energy. The coincidence limit of the scattering Green tensor is well defined for source and field points approaching the surface from the vacuum side, and the Casimir--Polder potential follows directly from Eq.~\eqref{eq:UCP} with the scattering part of the planar Green tensor~\cite{10.1063/5.0106503}.
Taking the limit $z\to0^+$ yields a finite contact value,
\begin{equation}
U_{\rm CP}(0)
=
\frac{\hbar\mu_0}{2\pi}
\int\limits_0^\infty \! \mathrm d\xi\,
\xi^2\,\alpha(\mi\xi)\,
\mathrm{Tr}\,\mathbf G^{(1)}_{\rm planar}(z=0,\mi\xi)\,,
\end{equation}
which coincides with the Casimir--Polder energy of an atom located directly at a planar interface in vacuum. For an isotropic atom, this can be written explicitly as
\begin{align}
U^{\rm V}_{\rm CP}(0)
=
-\frac{\hbar}{8\pi^2\varepsilon_0}
\int\limits_0^\infty \mathrm d\xi\,\alpha(\mi\xi)
\int\limits_0^\infty \mathrm d k_\parallel k_\parallel
\left[
r_{\rm s}(\mi\xi,k_\parallel)
+
r_{\rm p}(\mi\xi,k_\parallel)
\right]\,,
\end{align}
where $r_{\rm s,p}$ are the Fresnel reflection coefficients of the planar interface, leading to a surface-modified and together with Eq.~\eqref{eq:UCPI+II} a medium-assisted-surface-modified Lamb shift.

\begin{figure}
    \centering
    \includegraphics[width=0.6\columnwidth]{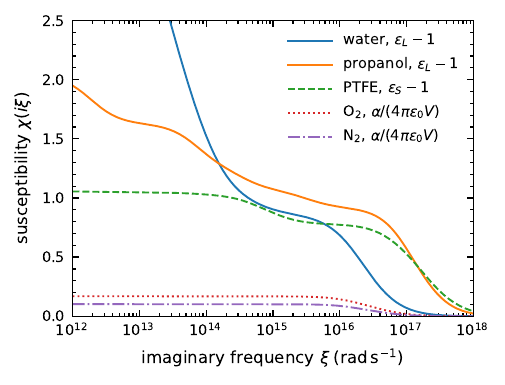}
    \caption{Susceptibilities of water (solid blue line) from Ref.~\cite{water}, propanol (solid orange line) and polytetrafluoroethylene (PTFE) (dashed green line) from Ref.~\cite{PhysRevA.81.062502} and the normalised polarisabilities for oxygen (dotted red line) and nitrogen (dash-dotted purple line) from Ref.~\cite{doi:10.1021/acs.jpca.7b10159}.}
    \label{fig:dielectrics}
\end{figure}

\section{Numerical Examples}
In this section, we illustrate the general theory derived above using four representative molecular--liquid--surface combinations.  
We consider O$_2$ and N$_2$ molecules dissolved in either water or propanol interacting with a planar polytetrafluoroethylene (PTFE) interface.
These systems span a broad range of dielectric responses—from strongly polar water to weakly polar propanol—and therefore constitute suitable test cases for examining the sensitivity of the Onsager real-cavity model to the surrounding medium.

Figure~\ref{fig:dielectrics} shows the relevant susceptibilities $\chi(\mi\xi) = \varepsilon(\mi\xi)-1$ for water, propanol, and PTFE, together with the normalised molecular susceptibilities $\alpha(\mi\xi)/(4\pi\varepsilon_0 V)$ of O$_2$ and N$_2$. The data are taken from Refs.~\cite{water,PhysRevA.81.062502,doi:10.1021/acs.jpca.7b10159}. These spectral functions enter the Casimir--Polder interaction solely through the frequency-dependent material response, and thus Fig.~\ref{fig:dielectrics} provides direct insight into the relative weight of Matsubara frequencies contributing to the dispersion interaction.

From these susceptibilities we compute the medium-assisted coefficients $C_3^{\rm med}$, the liquid-screened coefficients $C_3^{\rm L}$, and the vacuum--surface coefficients $C_3^{\rm S}$ according to Eqs.~(\ref{eq:C3free}) and~(\ref{eq:UPCC3}).  
The resulting values are summarised in Table~\ref{tab:parameters}. As expected, the strong permittivity of water yields significantly larger medium-assisted coefficients than propanol, while the differences between O$_2$ and N$_2$ reflect their molecular volumes and polarisability spectra primarily.

\begin{table}[htb]
    \centering
    \begin{tabular}{c|c|c|c}
         & Water & Propanol & \multirow{2}{*}{PTFE} \\
         & $R_2 =1.68\,\text{\AA}$ & $R_2=3.497\,\rm{\AA}$ \\\hline
         O$_2$ & $C_3 = 0.565$ & $C_3 = 1.085$ & \multirow{2}{*}{$C_3 = 0.983$} \\
         $R_1 = 2.187\,\text{\AA}$ & $\tilde{C}_3 = 0.406$ & $\tilde{C}_3 = 0.081 $\\\hline
         N$_2$ & $C_3 = 0.428$ & $C_3 = 0.806$ & \multirow{2}{*}{$C_3 = 0.730$} \\
         $R_1 = 2.206\,\text{\AA}$ & $\tilde{C}_3 = 0.294$ & $\tilde{C}_3 = 0.061 $ \\
    \end{tabular}
    \caption{Overview of the $C_3$-coefficients ($\rm{meV(nm)}^{3}$) in free-space $C_3$, Eq.~(\ref{eq:C3free}), and medium-assisted $\tilde{C}_3$, Eq.~(\ref{eq:UPCC3}), for the material combinations: oxygen and nitrogen molecules dissolved in water and propanol in front of a PTFE interface with the considered cavity radii $R_1$ and van der Waals radii $R_2$. The polarisability data and cavity radii are taken from Ref.~\cite{doi:10.1021/acs.jpca.7b10159}, the dielectric function of water from Ref.~\cite{water} and of propanol and PTFE from Ref.~\cite{PhysRevA.81.062502}, and the van der Waals radii from Ref.~\cite{Properties_Reid} via the van der Waals constant $b$, $R_2 =\sqrt[3]{3b/4\pi N_{\rm A}}$ with the Avogadro constant $N_{\rm A}$.}
    \label{tab:parameters}
\end{table}

\begin{figure}
    \centering
    \includegraphics[width=0.6\linewidth]{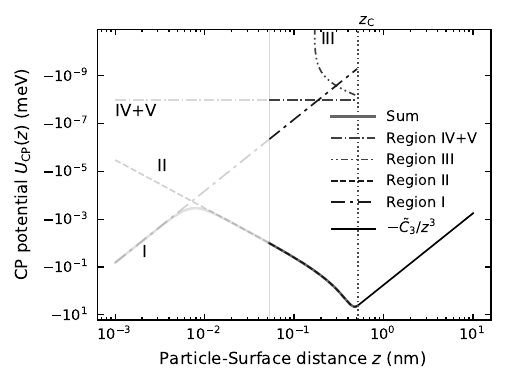}
    \caption{Decomposition of the non-retarded Casimir--Polder potential for an O$_2$ molecule in water near a PTFE interface within the open-cavity model. The total potential (solid line) is shown together with the individual geometric contributions from Regions IV+V (dashed-dotted line), III (dashed-double-dotted line), II (dashed line), and I (long dashed-dotted line). The vertical dotted line marks the crossover position $z_{\rm C}$, beyond which the potential is continued by the asymptotic $-\tilde{C}_3/z^3$ form (solid line). Regions III–V are negligible over the entire physically relevant range ($\ge a_0 \approx 0.0529\, \rm{nm}$; vertical line).
    Region I corresponds to direct molecule–surface attraction and becomes significant only after screening by Region II is overcome in the irrelevant regime. Thus, the dominant contribution responsible for the transient sign change of the force originates from Region II.}
    \label{fig:pot_contirbutions}
\end{figure}

Figure~\ref{fig:pot_contirbutions} illustrates the decomposition of the non-retarded Casimir--Polder potential for oxygen in water near a PTFE surface into its geometrical contributions within the open-cavity construction. The total potential (solid line) exhibits a non-monotonic behaviour with a pronounced maximum close to the crossover distance $z_{\rm C}$. The decomposition reveals that Regions III, IV, and V do not contribute in any appreciable manner over the physically relevant distance range. In particular, the contribution arising from Region III remains several orders of magnitude below the dominant terms and therefore cannot influence the qualitative structure of the interaction.

Region I corresponds to the direct molecule–surface interaction in the absence of screening. As expected, this contribution is purely attractive and follows the characteristic $z^{-3}$ scaling at larger distances. However, it becomes relevant only once the dielectric screening mediated by Region II is sufficiently reduced, which only occurs in the sub-atomic regime for distances smaller than the Bohr radius 
($z\le a_0 \approx 0.0529\, \rm nm$).
In the distance range of practical interest (sub-nanometre to few-nanometre regime), Region I alone does not determine the interaction.

The physically decisive contribution stems from Region II. This term captures the modification of the field inside the cavity due to the dielectric contrast across the interface. Its magnitude temporarily exceeds the screened molecule–surface attraction, leading to a localised inversion of the force in the contact region. The non-monotonic structure of the total potential can therefore be traced unambiguously to the interplay between Region II and the asymptotic surface attraction. The dominance of Region II demonstrates that the effect is governed by macroscopic dielectric response rather than microscopic configurational details, supporting the use of continuum-based descriptions in this parameter regime.

Beyond the crossover position $z_{\rm C}$, the full result smoothly approaches the asymptotic continuation $-\tilde{C}_3/z^3$, confirming the internal consistency of the construction.

\begin{figure}
    \centering
    \includegraphics[width=0.6\linewidth]{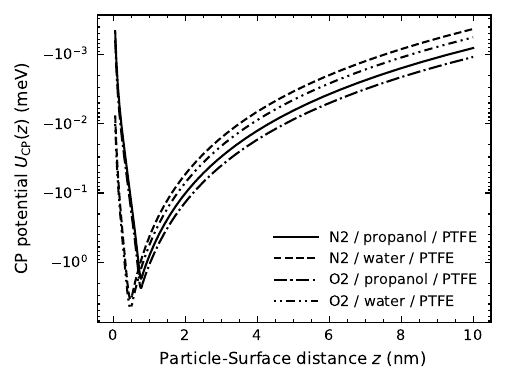}
    \caption{Total nonretarded Casimir–Polder potentials for O$_2$ and N$_2$ molecules in water and propanol near a PTFE interface. For each case, the open-cavity result is smoothly connected to its asymptotic $-\tilde{C}_3/z^3$ continuation beyond $z_{\rm C}$. The qualitative behaviour is robust across all combinations, with quantitative differences reflecting molecular polarisability and solvent dielectric screening.}
    \label{fig:UCP}
\end{figure}

The total non-retarded Casimir--Polder potentials for all investigated combinations are shown in Fig.~\ref{fig:UCP}. Despite quantitative differences in magnitude, the qualitative structure of the interaction remains remarkably robust across all systems. In particular, the characteristic crossover behaviour and the transient maximum near the cavity–interface transition persist for both molecular species and solvents.

The overall scale of the interaction is primarily governed by the molecular polarisability. As expected, oxygen exhibits systematically larger interaction energies than nitrogen, reflecting its larger dynamic polarisability. This trend holds independently of the surrounding liquid.

The solvent dependence enters through dielectric screening. Comparing water and propanol, the higher permittivity of water enhances screening of the direct molecule–surface interaction, leading to a reduction in the asymptotic tail. Nevertheless, the non-monotonic feature near the crossover distance remains present in both media, indicating that it is not a consequence of a particular dielectric function but rather an intrinsic geometric effect of the open-cavity construction at an interface.

Importantly, the continuation to the asymptotic $-\tilde{C}_3/z^3$ form occurs smoothly in all cases, confirming that the cavity-modified interaction consistently connects to the standard planar limit at larger distances. The persistence of the intermediate force inversion across all molecular and solvent combinations demonstrates that the effect is not material-specific but arises generically from the interplay of dielectric screening and cavity geometry.

\section{Conclusions}
We have developed and applied an extended Onsager real-cavity approach to describe the Casimir--Polder interaction of small molecules dissolved in dielectric liquids near planar interfaces. By resolving the geometry of the cavity opening, we derived analytic expressions for the five distinct regions contributing to the interaction and demonstrated how they connect smoothly to the asymptotic medium-assisted power law at large particle--surface separations. Using experimentally established dielectric response functions for water, propanol, and PTFE, together with accurate molecular polarisabilities for O$_2$ and N$_2$, we evaluated the full distance-dependent interaction for representative molecular--liquid--surface combinations.

Beyond providing quantitative estimates, the present framework offers a transparent analytical decomposition of the interaction into well-defined geometric and material contributions. This enables a direct identification of the dominant physical mechanisms across different regimes and allows for rapid exploration of parameter dependencies, such as solvent permittivity, molecular polarisability, and cavity geometry. In contrast to purely numerical approaches, the analytic structure enables disentangling local-field screening, cavity-opening effects, and surface contributions, thereby providing a useful baseline for interpreting and guiding more detailed microscopic calculations.

The resulting interaction reflects the interplay between local-field screening within the cavity and geometry-dependent modifications at the interface, leading to distinct behaviour across different liquids and molecular species. The model consistently reproduces the expected $-C_3/z^3$ asymptotics at large separations and captures how deviations from a closed cavity modify the interaction as the molecule approaches the interface.

The present description is based on a continuum dielectric response and a dipole approximation for the molecule. It is therefore expected to be quantitatively reliable once the particle--surface separation exceeds a few molecular or lattice spacings, whereas at smaller separations, microscopic structure, electronic overlap, and short-range exchange interactions become increasingly important. In this regime, the results should be interpreted as a controlled extrapolation of the continuum model.

This limitation becomes particularly relevant in hydrogen-bonding liquids such as water, where local solvent structuring and hydration-shell formation introduce molecular-scale correlations that are difficult to capture within a purely continuum dielectric description. 
Similarly, polar molecules may exhibit additional interaction channels associated with permanent dipoles and orientational effects beyond the purely dispersive contribution considered here. Nevertheless, the underlying Green-function framework is not restricted to dispersion interactions alone. As discussed in Ref.~\cite{FiedlerUnifiedWeakInteractions}, weak intermolecular interactions can be formulated within a unified Green-function description in complex electromagnetic environments. In the present geometry, reinserting the definition of the $C_3$ coefficient into the final expressions yields the corresponding approximate Green tensor, which may subsequently be employed in the static limit to analyse additional dipolar interaction channels within the same formalism.

Extensions of the present framework could incorporate finite-size and multipole corrections, dielectric environments that allow for sign reversals of the interaction, and a systematic coupling to microscopic models of the cavity region. Previous work has shown that local-field effects associated with finite-size molecules may give rise to repulsive contributions at larger separations~\cite{doi:10.1021/acs.jpca.7b10159}, suggesting that combining cavity geometry with more refined molecular response models may provide a route toward describing a broader class of dispersion interactions in complex environments.

Overall, the present theory provides a consistent and analytically tractable framework for understanding how local-field effects, molecular response, and cavity geometry jointly shape Casimir--Polder interactions in liquids, and offers a complementary perspective to fully microscopic approaches.

While the quantitative magnitude of the local-field corrections depends on the specific cavity construction, the overall physical picture obtained here is expected to be robust. In particular, the crossover between bulk-like screening and interface-modified open-cavity behaviour arises from the generic breaking of spherical symmetry near the surface, rather than from the precise geometry of the cavity opening. The present geometry was chosen as the simplest analytically tractable model that continuously interpolates between the closed-cavity and fully surface-exposed limits.

\appendix
\section{Long expressions for Region IV}\label{app:long}
In the cavity-opening region IV, the potential reads
\begin{equation}
    U_{\rm CP}^{\rm IV}(z) = -\frac{12 C_3^{\rm L}R_2}{\pi R_1^3\left(R_1 + 2R_2\right)^3}\frac{\Lambda_{\rm N}}{\Lambda_{\rm D}}\,.
\end{equation}
Here
\begingroup
\allowdisplaybreaks
\begin{align}
    \lefteqn{\Lambda_{\rm N} = 8 R_2^7\tilde{h} + 64 R_2^6\tilde{h}\left(R_1 + \frac{3}{8}h_3 + \frac{z}{2}\right)+ R_2^5\left\lbrace 12R_1^2 \left(9\tilde{h} - 1\right) + 4R_1 \left[h_3\left(14\tilde{h} - 1\right) - 3z\right]\right.}\nonumber\\
    &\left.- 16\tilde{h} z\left(h_3 + 3 z\right)\right\rbrace
    +R_2^4 \left\lbrace 2R_1^3\left(50\tilde{h} - 11\right) + R_1^2\left[64 h_3\left(\tilde{h} - \frac{5}{32}\right) -2 z \left(16\tilde{h} - 7\right)\right]\right.\nonumber\\
    &\left.- 32 zR_1\left[z\left(\frac{3}{2}\tilde{h} - \frac{1}{4}\right) + h_3\tilde{h}\right]+ 16\tilde{h} {z}^3\right\rbrace + 58R_1R_2^3 
   \left\lbrace
   R_1^3\left(\tilde{h} - \frac{9}{29}\right) \right.\nonumber\\
    &\left. + R_1^2\left[-\frac{z}{29} \left(16\tilde{h} +3\right) + \frac{20}{29} h_3\left(\tilde{h} - \frac{1}{5}\right)\right]- \frac{16}{29} R_1 z \left[ z\left(\frac{3}{4}\tilde{h} - \frac{3}{8}\right) + h_3\tilde{h}\right]+ \frac{8}{29}{z}^3\tilde{h}
    \right\rbrace\nonumber\\
    &+ 24 R_1^2 R_2^2 \left\lbrace R_1^3\left(\tilde{h}- \frac{7}{24}\right) + R_1^2\left[\frac{7}{12}h_3\left(\tilde{h} - \frac{1}{7}\right) -z\left(\frac{\tilde{h}}{3}  - \frac{1}{24}\right)  \right]  - \frac{2 z}{3}\left( R_1h_3\tilde{h} -\frac{z}{2} \right) + \frac{\tilde{h}{z}^3}{3}\right\rbrace \nonumber\\
    &+ 7 R_1^4 R_2\left[R_1^2\left(\tilde{h} - \frac{1}{7}\right) + \frac{R_1}{7}\left(2 h_3\tilde{h}+ z\right) - \frac{4z}{7}\left(h_3\tilde{h}-\frac{z}{2}\right)\right] + R_1^7\tilde{h}\,,
    \end{align}
    and
    \begin{align}
    \lefteqn{\Lambda_{\rm D} =5R_1^4R_2 + 12R_1^3R_2^2 + 8R_1^2R_2^2\left(3 R_2 -z\right)+ 2 R_2 h_3 \left(\left[R_1+ R_2\right]^2 + R_2^2\right)\left(R_1 + 3R_2 - 2z\right)}\nonumber\\
    &+ R_1^5  + 4R_1R_2^3\left(7R_2 - 4z\right)+ 4 R_2^2\left(R_2^3 + 4R_2^2z - 6R_2{z}^2 + 2{z}^3\right)\,,\qquad\qquad \qquad
    \end{align}
    where
    \begin{align}
    \tan h_1 &= \frac{2R_2( R_2 - R_1-2z)h_3 + \left[\left(R_1+R_2\right)^2+ R_2^2\right](R_1 + z)}{R_1(R_1 + 2R_2)h_3}\,,\\
    h_2 &= \arctan\frac{2R_2(R_2-z )}{R_1 (R_1 + 2 R_2)}\,,\\
    h_3 &= \sqrt{(R_1 + z)(R_1 + 2R_2 - z)}\,,\\
    \tilde{h} &= h_1-h_2\,.
\end{align}
\endgroup

\bibliographystyle{unsrt}  
\bibliography{references}

\end{document}